\documentclass[fleqn,12pt,twoside]{article}
\usepackage{espcrc1}
\usepackage{amssymb}

\usepackage{graphicx}

\newcommand{\AmS}{{\protect\the\textfont2
  A\kern-.1667em\lower.5ex\hbox{M}\kern-.125emS}}
\newcommand{\dd}{{\textrm d}}
\def\hs{\hat{s}}
\def\htm{\hat{t}}
\def\hu{\hat{u}}

\title{Multiple scattering and $p_t$-broadening at RHIC energies}

\author{G. Papp\address[ELTE]{HAS Research Group for Theoretical Physics,
E\"otv\"os University, Budapest, H-1518 Hungary}\address[KSU]{CNR, 
Department of Physics, Kent State University, Kent
OH 44242, USA}, G.G. Barnaf\"oldi\address[KFKI]{RMKI Research Institute for
Particle and Nuclear Physics, Budapest, Hungary}, 
G. Fai\addressmark[KSU], P. L\'evai\addressmark[KFKI] and 
Y. Zhang\addressmark[KSU]}

\begin{document}

\maketitle

\begin{abstract}
In ultrarelativistic heavy-ion collisions, in the 2 GeV$<p_\perp<$ 6 GeV 
transverse momentum region, the soft and semi-hard multiple scattering of the
incoming nucleons in the nuclear medium 
results in the broadening of the expected hadronic
(e.g. pion) $p_\perp$ spectra relative to proton-proton ($pp$) collisions. 
Thus, higher transverse-momentum regions are 
populated than in $pp$ collisions. In a perturbative QCD
based calculation we include the intrinsic transverse momentum ($k_\perp$)
of the partons in the nucleon (determined from $pp$ collisions), augmented 
by the extra broadening obtained via a systematic analysis of proton-nucleus 
($pA$) collisions
in the energy range 17$<\sqrt{s}<$ 39 AGeV. The original polynomial
spectra are modified, and a nearly exponential spectrum appears in the
region
2$\lesssim p_\perp\lesssim 3.5$ GeV. At present RHIC 
energies ($\sqrt{s}=$130 AGeV), the
slope of the calculated spectra is reminiscent of that of fluid-dynamical 
descriptions, but
lacks any thermal origin. We determine and discuss the size of the 
modifications originating in multiple scattering, which lead to this state
of affairs.
\end{abstract}

\section{Introduction}

With the start of RHIC operation~\cite{QM2001} new data are now
available for high energy heavy ion collisions.
This makes it possible to contrast the models to the experimental data
and to learn about new effects in this so far unreachable domain.

In this talk we discuss the features of some hadronic spectra in
Au+Au collisions at 130 GeV. First we give an introduction to the
pQCD-improved parton model and fix our parameters in $pp$
collisions. Next, we study the available $pA$ experiments
in order to understand the nuclear (Cronin) effect, the enhancement
of particle production in a certain $p_\perp$ range compared
to $pp$ collisons. Finally, we present our calculation for 
nucleus-nucleus ($AA$) collisions at present RHIC energy.

\section{Perturbative QCD}

The classical pQCD parton model assumes factorization of the hadron
(meson $M$) production
cross section, namely, the particle production is interpreted as
a superposition of elementary, partonic collisions~\cite{XN},
\begin{equation}
\label{eq-kt0}
  E_M\frac{\dd\sigma_M^{pp}}{\dd^3p} =
        \sum_{abcd}\!  \int\!\dd x_a \dd x_b\dd z_c\ f_{a/p}(x_a,Q^2)\
        f_{b/p}(x_b,Q^2) \frac{\dd\sigma_{ab\to cd}}{\dd\htm}\,
		\frac{D_{M/c}(z_c,\!{\hat Q}^2)}{\pi z_c^2}
		\hs \delta(\hs\!+\!\htm\!+\!\hu),
\end{equation}
where $f_{a/p}(x_a,Q^2)$ is the parton distribution function (PDF), i.e.
the probability of finding parton
$a$ in the proton with momentum fraction $x_a$ at transverse momentum
scale $Q$, $d\sigma/d\htm$ is the pQCD calculable cross section of quarks and
gluons, while $D_{M/c}(z_c,\!{\hat Q}^2)$ is the fragmentation
function (FF) of parton $c$ into a meson $M$ with the meson carrying 
a momentum fraction $z_c$ of the parton at transverse momentum scale
$\hat Q$. The PDF and FF, contain all the non-perturbative
information and are fixed by experimental data. In our calculation
we use the leading order GRV~\cite{GRV} PDF and the leading order
KKP~\cite{KKP} FF. (We also carried out approximate NLO 
calculations~\cite{NLOus}).

\begin{figure}[htb]
\begin{minipage}[t]{75mm}
\includegraphics[height=75mm]{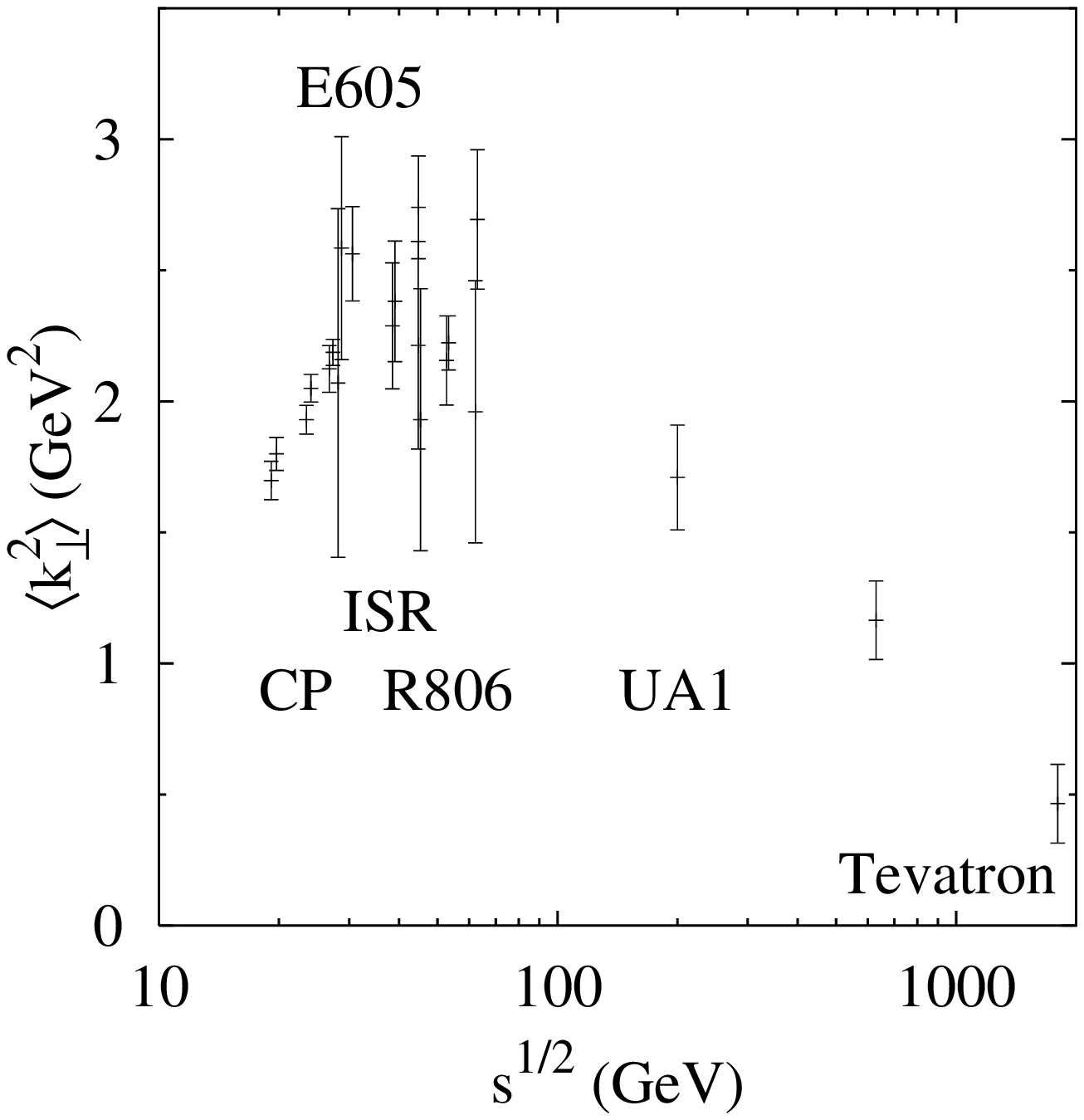}
\caption{Average transverse momentum square fitted to $pp$ experiments}
\label{fig-kt}
\end{minipage}
\hspace{\fill}
\begin{minipage}[t]{75mm}
\includegraphics[height=75mm]{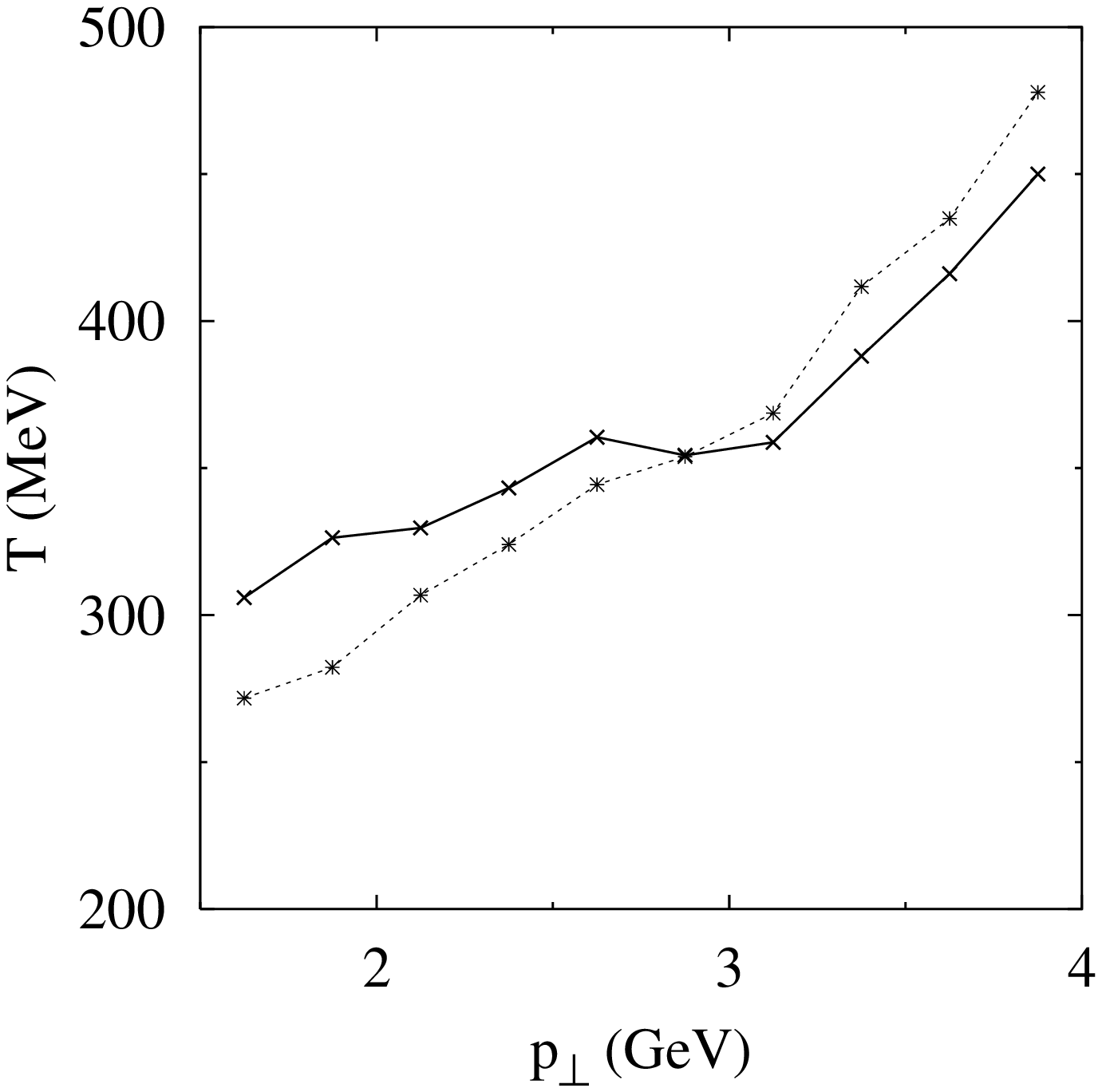}
\caption{Slope parameter $T$, for scaled $pp$ collision (dotted line) and
central Au+Au collisions at 130 GeV (solid line) in $\pi$ production.}
\label{fig-T}
\end{minipage}
\end{figure}

Evaluating integral~(\ref{eq-kt0}) numerically and comparing it to existing
experimental data of charged and neutral pion and kaon production in 
$pp$ and $p\bar{p}$ reactions in the range of $\sqrt{s}=17-1800$ GeV we
found a factor of two discrepancy with the theory underestimating the
yield. Even more sophisticated NLO calculations failed
to reproduce the pion spectra~\cite{Aurenche}. This motivates us to
introduce another non-perturbative parameter, the intrinsic transverse
momentum of partons. Such a quantity is not fictitious: it has been
measured experimentally~\cite{Corcoran}. We consider the distribution
of this intrinsic transverse momentum to be Gaussian for simplicity,
and fix its width from experiments. The integral for both incoming
partons in Eq.~(\ref{eq-kt0}) is modified to
\begin{equation}
\label{eq-kt}
\dd x_a \ f_{a/p}(x_a,Q^2) \longrightarrow \dd x_a
\ \dd^2\!k_{\perp,a}\ \frac{e^{-k_{\perp,a}^2/\langle k_{\perp}^2 \rangle}}
{\pi \langle k_{\perp}^2 \rangle} \  f_{a/p}(x_a,Q^2).
\end{equation}
The fit of the intrinsic transverse momentum width is shown in
Fig.~\ref{fig-kt} for pions. A similar plot may be obtained for
kaons (with much less experimental data available). Using the
intrinsic transverse momentum distribution with a width given by the figure
we were able to fit $pp$ experimental data within 30\% for
$p_\perp>2$ GeV.

Having fitted the $pp$ data using Eqs.~(\ref{eq-kt0}-\ref{eq-kt}) 
parameterized by the $\langle k_\perp^2\rangle_{pp}$ values, 
we now turn to the $pA$ collisions. 
It was found in experiments that there is an extra nuclear enhancement
in $pA$ collisions compared to the simple scaling of $pp$ data in a
certain $p_\perp$ range (Cronin effect)~\cite{Cronin}. One needs to 
extend integral~(\ref{eq-kt0}-\ref{eq-kt}) as
\begin{equation}
\label{eq-pA}
  E_M\frac{\dd\sigma_M^{pA}}{\dd^3p} =
        {\int \dd^2b\ t_A(b)}\ E_M\frac{\dd\sigma_M^{nn}
  (\langle k_\perp^2\rangle_{pA},\langle k_\perp^2\rangle_{pp})}{\dd^3p}.
\end{equation}
Here $t_A(b)$ is the nuclear thickness function at impact parameter
$b$, $t_A(b)=\int\!dz \varrho(b,z)$ and the nucleon-nucleon cross section has
a dependence on the width of intrinsic transverse momentum of the proton
and nucleons in nuclei $A$, respectively. 
Furthermore, due to the nuclear
environment the partonic distribution functions are modified with a
shadowing and antishadowing region~\cite{shadow}.

The theoretical interpretation of the Cronin effect is related to the
nuclear environment, the average intrinsic transverse momentum of a
parton increases due to multiscattering,
\begin{equation}
\langle k_\perp^2\rangle_{pA} = \langle k_\perp^2\rangle_{pp} + C\ 
	h(\nu_A(b) - 1),
\end{equation}
where $\nu_A(b)$ is the number of target nucleons in the channel swept
by the incoming proton at impact parameter $b$, $h$ is an ``effectivity''
function, selecting the number of target nucleons contributing to the
$\langle k_\perp^2\rangle$ enhancement of the projectile and $C$ is the
average momentum square imparted in one nucleon-nucleon collision.

For the effectivity function we tried a function linearly rising with the
number of particles inside the channel and saturating after 
$n$ collisions. Requiring approximate target and energy independence 
of $h$ we found that the best fit is achieved using $n=3-4$ with average
momentum
square impart $C_{n=4}\approx 0.4$ GeV$^2$. In the following we 
use this value
for central nucleus-nucleus collisions.

Next, we performed a simulation of the ongoing RHIC experiments on meson
production at $\sqrt{s}=$130 GeV in Au+Au collisions. This requires a
modification of Eq.~(\ref{eq-pA}) extending it to the case of two colliding
nuclei,
\begin{equation}
\label{eq-AA}
  E_M\frac{\dd\sigma_M^{AB}}{\dd^3p} =
        {\int\!\dd^2b\,\dd^2r \ t_A(b)t_B(\vec{b}-\vec{r})}\ 
	E_M\frac{\dd\sigma_M^{nn}
  (\langle k_\perp^2\rangle_{pA},\langle k_\perp^2\rangle_{pB})}{\dd^3p}.
\end{equation}
Based on Fig.~\ref{fig-kt} we used $\langle k_\perp^2\rangle_{pp}=1.6$
GeV$^2$. A simple calculation without the nuclear enhancement and shadowing 
agrees well with the peripheral PHENIX results on $\pi^0$ 
production~\cite{Gabor}. However, for central collisions there is a
disagreement between this pQCD model and the experimental data, indicating 
a strong suppression in the particle production. A possible candidate
explaining this phenomenon is jet quenching~\cite{Quench}.

An interesting quantity to study is the slope of the meson production
cross section. For peripheral collisions (no shadowing and nuclear 
enhancement) a pQCD calculation yields a non-exponential spectrum with
a slope parameter rising with the transverse momentum (Fig.~\ref{fig-T}
dotted line). However, in case of central collisions in a $p_\perp$
window 2--3.5 GeV the slope parameter is almost constant resembling
an exponential (thermal) spectrum. Predictions for lower transverse
momentum are uncontrollable within pQCD. Our studies show that at
SPS energies this plateau is more pronounced and extends to the highest
transverse momenta measured.

\section{Conclusion}
In this talk we showed that at high transverse momenta ($p_\perp>2$ GeV) 
pQCD augmented with an intrinsic transverse momentum distribution of
the partons gives a good description of $pp$, $pA$ and $AA$ collisions.
A ``thermal'' slope is observed in central $AA$ collisions in a 
transverse momentum window 2--3.5 GeV despite the non-thermal
character of the model.
At RHIC energies, however, a strong suppression is observed in
hadron production in central collisions compared to pQCD calculations.
A possible explanation is presented in~\cite{Quench}.

\section*{Acknowledgments} 
This work was supported by US DOE grant under DE-FG02-86ER40251, 
Hungarian grants OTKA T032796, T034842, FKFP 220/2000 and by the MTA-OTKA-NSF 
grant INT-0000211 (MTA-038 OTKA-33423). The authors
acknowledge the computing support of BCPL (Bergen,Norway).

\end{document}